\documentclass[journal]{new-aiaa}
\usepackage[utf8]{inputenc}

\usepackage{graphicx}
\usepackage{amsmath}
\usepackage[version=4]{mhchem}
\usepackage{siunitx}
\usepackage{longtable,tabularx}
\usepackage[utf8]{inputenc}

\usepackage{graphicx}
\usepackage{amsmath}
\usepackage[version=4]{mhchem}
\usepackage{siunitx}
\usepackage{longtable,tabularx}
\usepackage[utf8]{inputenc}

\usepackage{subcaption}
\usepackage{graphicx}
\usepackage{amsmath}
\usepackage[version=4]{mhchem}
\usepackage{longtable,tabularx}
\usepackage{float}
\usepackage{siunitx}
\usepackage{datetime2}
\usepackage{etoc}
\usepackage{todonotes}
\title{Implementation of a Low-Cost Flight Tracking System for High-Altitude Ballooning}

\author{Caitlyn A. K. Singam\footnote{Student, University of Maryland, College Park}}
\affil{University of Maryland, College Park, College Park, MD, 20742}

\begin{document}

\maketitle

\begin{abstract}
High altitude balloons (HABs) are typically tracked via GPS data sent via real-time radio-based communication systems such as the Automated Packet Reporting System (APRS). Prefabricated APRS-compatible tracker modules have made it trivial to transmit GPS coordinates and payload parameters in compliance with the requisite AX.25 protocol. However, in order to receive and track APRS signals, conventional methodologies call for the use of a Very High Frequency (VHF) receiver to demodulate signals transmitted on the 440/144 MHz APRS frequencies, along with a compatible antenna and custom methodology for visualizing the HAB's location on a map. The entire assembly is typically costly, cumbersome, and may require an internet connection in order to obtain real-time visualization of the HAB's location. This paper describes a low-cost, handheld system based on open-source software that operates independently of an internet connection. The miniaturized system is suited to tracking done either from a vehicle or on foot, and is cost-effective enough to be within the means of nearly any HAB user. The paper also discusses preliminary test results and further applications. 
\end{abstract}

\section*{Nomenclature}

{\renewcommand\arraystretch{1.0}
\noindent\begin{longtable*}{@{}l @{\quad=\quad} l@{}}
HAB  & High Altitude Balloon \\
STEM & Science, Technology, Engineering, and Math \\
APRS& Automatic Packet Reporting System \\
GPS & Global Positioning System\\
ISM & Industrial, Scientific and Medical\\
STEM & Science, Technology, Engineering, and Math \\
YAAC &  Yet Another APRS Client \\

\end{longtable*}}

\section{Introduction}
\lettrine{H}{igh} altitude ballooning is an increasingly popular scientific platform for the investigation of atmospheric phenomena, aerial monitoring and sampling, and experimentation in a near-space environment. The relatively low cost associated with this platform has also made it appealing to institutions seeking to test prototypes of instruments with suborbital applications \cite{jones}, to educators as a hands-on instructional tool for STEM education \cite{voss}, and as a recreational activity. The recent discovery of microbiota in the upper troposphere \cite{deleon},\cite{yang} has also drawn interest from biologists \cite{singam2017designing} and other disciplines not traditionally affiliated with the aerospace field. 

A critical component of HAB is the recovery of payloads after a balloon burst or a controlled payload release. A biological payload, for example, relies on the quick recovery and analysis of samples that may degrade with time. Consequently, locating and recovering payloads in a timely manner and preserving the integrity of payloads is an essential component of scientific ballooning. However, locating a balloon is a non-trivial matter. During flight, a medium sized HAB (using a 1600 g latex balloon) can typically reach altitudes of 25 to 30 km before burst \cite{hall}, and remain airborne for a total of one to two hours depending on the amount of helium used. Although the high burst altitude and extended flight time are two of the more desirable characteristics of HABs, balloons will typically encounter high-velocity winds while airborne and can consequently drift several kilometers away from the launch site. Due to the unpredictability of individual air currents and of the balloon's precise movements, it is nearly impossible to predict a HAB's landing location to within less than a few kilometers of error. 

Furthermore, HAB payload strings are frequently prone to entanglement in brush, foliage, etc., which makes recovery solely by visually searching the predicted landing zone impractical. As a result, accurate positioning data is an essential requirement for high altitude ballooning and requires relatively accurate positioning data. 

Accurate positioning data is usually obtained by equipping the HAB with a GPS receiver. However, GPS data is of little value without the transmission of data from the airborne platform and the consequent access to the data via the ground tracking system.

\subsection{The Automatic Packet Reporting System}

APRS provides a convenient protocol which can be used to transmit data over amateur radio frequencies. The development of APRS is credited to Bob Bruninga \cite{bruninga}, who created the protocol as a "real-time local tactical communications system for rapidly exchanging digital data of immediate value to operations". Today, 144.39 MHz is dedicated throughout the North American continent for APRS use, 433.800 / 432.500 MHz in Europe, and 145.175 MHz in Australia.

Now more than two decades old, APRS distinguishes itself from conventional packet radio as a digital communications protocol for exchanging information between multiple stations covering the localized reception area. APRS specifies specific formats for time and position packets. The APRS protocol, which can be used to carry environmental and payload parameters such as wind speed, transmitter power, effective antenna height-above-average-terrain, antenna gain, antenna directivity etc. APRS is typically transmitted using the  the AX.25 Amateur Packet Radio Link Layer Protocol for Unnumbered Information (UI) Packets using 1200-bit/s Bell 202 AFSK. 

Since APRS operates on a frequency in the 2 meter/Very High Frequency (VHF) band, the signal from any individual station is mostly limited to line of sight transmissions over a relatively short range (typically a few kilometers at most if the transmitting station is operating at low power). A signal can be further dampened by thick vegetation or other obstructions that interrupt the line of sight between the transmitting and receiving stations, which is often an issue for downed balloons (which are often ensnared in trees or thickly vegetated areas). 

There is a network of fixed APRS digipeater stations around the globe that re-broadcast nearby APRS signals at higher power, thereby increasing the effective range of a HAB's APRS signal. Furthermore, other amateur operators with IGates receive and forward APRS signals to the APRS-Internet Service (APRS-IS) network \cite{loveall}, where they can be viewed from online tracking websites. However, viewing these sites requires a steady data connection, and the receipt of packets by the tracking team may be delayed depending on the speed of the wireless/cellular data connection at the tracking team's location.

\section{Conventional tracking systems}
\subsection{GPS acquisition and transmission}
With the increasing popularity of APRS, there has been a surge in the availability of commercially produced, pre-configured APRS-compatible transmitters. Two of the more commonly used APRS-compatible transmitter units are the Uputronics HABduino \cite{singam2017high} (produced in the United Kingdom) and the Byonics Micro-Trak. 

The HABduino may be purchased with a 300 mW Radiometrix HX1 144.390-10 transmitter for Automatic Packet Reporting System (APRS) transmission in the US and includes a 10 mW Radiometrix MTX2-433-10 for operation in the ISM band in the UK/Europe. GPS data is acquired via an onboard Ublox MAX8Q GPS designed for high altitude use and transmitted via the HX1 for reception by APRS networks on the ground \cite{uputronics}. The radios are equipped with compatible antennae prior to use - typically whip antennae in order to conserve on space and weight. 

Similarly, the Micro-Trak acquires GPS data once per second via a built-in GPS module tuned to the 1575.42 MHz frequency. It is set to use the Mic-Encoder (MIC-E) protocol by default, thereby circumventing the need for a TNC and enabling the transmission of shorter, higher throughput APRS AX.25 frames directly from the GPS unit \cite{byonics}.

\subsection{Receiver system}
The receiving station necessarily consists of hardware compatible with receiving data from the transmitters enabled on the balloon payloads. The traditional tracking system employed by one ballooning group \cite{umd}, for example, consists of a roof mounted omnidirectional Comet CA-2X-4SR Broadband VHF/UHF Dual Band Antenna that is designed to assist search and rescue volunteers and professionals. The dual-band antenna is VHF and/or UHF capable and may be used to simultaneous transmit and receive on both VHF and UHF. The high gain antenna which allows for an extended coverage area and receive range is typically attached to a Kenwood VHF/UHF FM Dual Band Data Communicator TM-D700A/E. It comes equipped with a built-in 1200/9600bps TNC compliant with AX.25 protocol, a D-sub 9-pin terminal for connection to a PC, a NMEA-0183 connector for GPS, and a remote panel with extra-large (188 x 54 pixel) backlit LCD and multifunction key display and extension cable \cite{kenwood}.

When tracking the balloon in terrain inaccessible to vehicles, some balloon groups have used handheld radios (e.g. Baofeng UV-5Rs) tuned to the APRS frequency and connected to an Android smartphone running APRSDroid. APRSDroid requires a TNC using a Bluetooth serial adapter or a direct wired connection for AFSK. The UV-5R radio is then tuned to the APRS frequency with its speaker output is connected to the microphone input of the smartphone\footnote{The description provided on page 10 of High Altitude Balloon Operations in the Mid-Atlantic States \cite{umd}, which states “The radio is then tuned to the APRS frequency and its mic output connected to the microphone input of the smartphone” is electrically unfeasible.}. The open source APRSDroid application running on the smartphone is then used to decode the packets for tracking the payload on foot. The connecting cable is generally an APRS-K2 TRRS cable that adapts the accessory jack on the BaoFeng (or similar) radio to a 3.5 mm TRRS jack on the phone. The APRS-K2 cable uses the virtual TNC found in APRSDroid to provide tracking information on the smartphone. It is important to ensure that the speaker output from the radio does not overload the smartphone audio circuits. 

\subsubsection{Electrical characteristics of the Android audio jack}

The Android audio jack is a 4 conductor 3.5 mm TRRS plug that follows CTIA specifications .  The typical pin configuration is illustrated in Figure 1a.

\subsubsection{Electrical characteristics of the BaoFeng accessory jack}
The BaoFeng accessory pinout is well document in associated user manuals.  The illustration in Figure 1b shows the typical configuration, for example, applicable to the BaoFeng F8HP.

\begin{figure}
	\begin{subfigure}[b]{0.45\textwidth}
		\includegraphics[width=\linewidth]{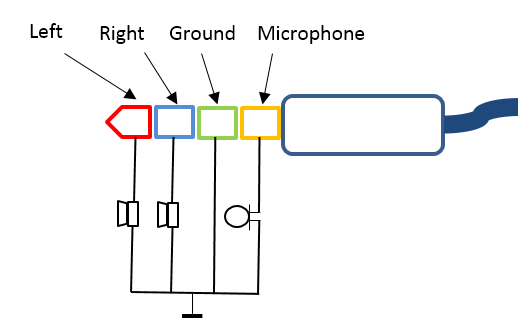}
		\caption{A diagram of the Android accessory jack.}
		\label{log}
	\end{subfigure}
	\hfill
	\begin{subfigure}[b]{0.45\textwidth}
		\includegraphics[width=\linewidth]{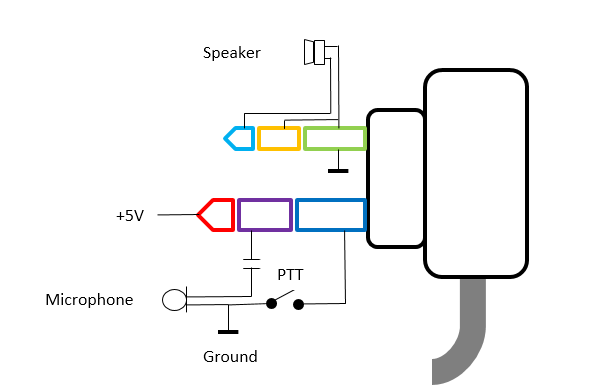}
		\caption{A diagram of the BaoFeng accessory jack.}
		\label{ny}
	\end{subfigure}
    \hfill
    \caption{Characteristics of the Android and BaoFend audio jacks.}
    \label{fig0}
\end{figure}

\subsubsection{The wired connection}
The connecting cable is generally an APRS-K2 TRRS cable  that adapts the accessory jack on the BaoFeng (or similar) radio to a 3.5 mm TRRS jack on the phone.  The APRS-K2 cable uses the virtual TNC found in APRSDroid to provide tracking information on the smartphone.  It is important to ensure that the speaker output from the radio does not overload the smartphone audio circuits.    

\subsection{The need for a simpler, more versatile tracking system}

Although APRS is a highly reliable radio communications system, it is limited in its usefulness once the HAB disappears below the tree line upon landing. Repeater stations and IGates are only helpful when the HAB is airborne and can maintain line-of-sight; once the balloon disappears below the treeline, it is often extremely difficult for a distant or fixed-location receiver station to maintain radio contact due to the presence of foliage, etc. that impedes long-distance signal transmission. When this occurs, the best means of re-establishing communications is by setting out on foot and decreasing the amount of distance (and foliage/obstacles) between the balloon and the individual tracking the balloon.

Many groups choose to instead supplement their APRS-based tracking systems with cellular tracker systems such as the Iridium system, which charges users a subscription fee to periodically send out text messages with the HAB's final coordinates after the balloon lands. However, the main disadvantage of this system is that cellular service reception is sometimes intermittent, or at worst nearly nonexistent, in many rural and remote locations. Furthermore, the lack of cellular reception at HAB landing sites can also compromise smartphone-based mobile tracking systems that require a cellular data connection in order to maintain access to a live-updating map showing the HAB's position. 

Issues with cellular reception are hardly rare in the high altitude ballooning community, as airspace restrictions (particularly in urbanized areas) and the safety issues associated with landing in a residential or commercial district tend to motivate HAB groups to launch in rural areas, which tend to have fewer cell towers. 

Therefore, there is a need for a new alternative to conventional tracking systems that is both versatile enough to facilitate on-foot tracking yet independent from a cellular data connection are. This paper demonstrates, completely open source, more agile, more versatile and less costly options that can serve as a dedicated tracking system.

\section{An alternative radio system for all-terrain tracking}

The proposed system, VIPER (the Vehiculated Instrument for Parameter Enquiry and Reporting), is based entirely on low-cost, open-source parts and software that are well within the means of most HAB programs. 

\subsection{System hardware}

A BaoFeng F8HP high-power handheld radio with a Nagoya high gain 15.6” UHF / VHF whip antenna was used as the APRS receiver for the proof-of-concept design. The antenna may be switched for a Nagoya UT-308UV 21” Magnetic Mount VHF/UHF (144/430Mhz) Antenna with a SMA-Female connector, or another similar antenna, for use when traveling in a vehicle and for greater gain. Similarly, the Baofeng radio can be substituted with a similar or more powerful handheld radio. 

The radio is then connected via a generic USB audio dongle with volume control capabilities to a device capable of running Java-based applets and Arduino C. For the purposes of preliminary testing, a laptop was used, but the design was subsequently miniaturized to be compatible with a Raspberry Pi 3 Model B+, which is based on a 1.4GHz 64-bit quad-core processor, has a built-in dual-band wireless LAN, Bluetooth 4.2/BLE, and faster Ethernet. The miniaturized design employing the Raspberry Pi also utilized a 7" touchscreen display to circumvent the need for a mouse/keyboard to allow the user to interact with the microcontroller. 
In order to supply the GPS location, altitude, and angle of the VIPER module, and thereby allow for real-time comparison of a chase vehicle/recovery party's position in relation to that of the HAB, the controller unit (the laptop/Raspberry Pi) is connected to an Adafruit Ultimate GPS module and attached u.FL compatible rubber duck antenna. The initial prototype interfaced the GPS unit with a laptop via an Arduino Uno running a GPS NMEA \cite{langley1995nmea} sentence interpreter algorithm connected to USB A to USB B cable, while the miniaturized unit connected directly to the GPS unit via jumper wires. 

\subsection{System software}

VIPER utilizes the data it receives through APRS packets picked up by the handheld radio to ascertain the GPS location and (if provided) altitude of the HAB being tracked. It supplements this information with its own current location, provided by the Adafruit GPS module, and plots the location of both concurrently on a live-updating map. The Raspberry Pi that the miniaturized VIPER module uses runs Raspbian Stretch Lite \cite{raspi}, which is a minimal image based on Debian Stretch.

The decoding of the APRS soundbytes into a human-readable format is handled by an open-source "soundcard" AX.25 packet modem/TNC and APRS encoder/decoder software program, DireWolf \cite{direwolf}. The software functions by accepting audio input via a KISS TCP client application, whereupon the received packets can be interfaced with a secondary open-source software application, YAAC \cite{yaac}. 

YAAC was chosen based on its versatility and ease of customization. It was configured to read in APRS packets via AGWPE and GPS data from the Adafruit module via Serial port. Both data streams can be concurrently plotted and the software can be configured to allow for the display of track tails that allow for the display of a tracked object's previous locations over a requested time frame. 

\subsection{Preliminary test results}

An initial test of the VIPER system was conducted by merely verifying that the system – even with a relatively low-gain antenna – was capable of receiving APRS packets from a distance. The initial test, run in a vehicle with the handheld radio equipped with the aforementioned radio mounted whip, was successful and outputted a log of the raw ARPS packets received as shown in Figure 2a.  As evident from the outputted data, the system was able to extract information from packets with differing content, varying numbers of transmitted parameters, and from a diverse collection of different transmitter types. The data allowed for estimation of the range of the system via a brief comparison of the location data of the transmitter, when available, and the location of the author’s vehicle. In one instance, VIPER was able to successfully display the location of a high-power APRS transmitter located in New York while it was being tested in Maryland (Figure 2b; the author's position, indicated by the W3EAX callsign, is marked with an orange rectangle, and the New York station is marked with a dark blue rectangle.). 

\begin{figure}
	\begin{subfigure}[b]{0.45\textwidth}
		\includegraphics[width=\linewidth]{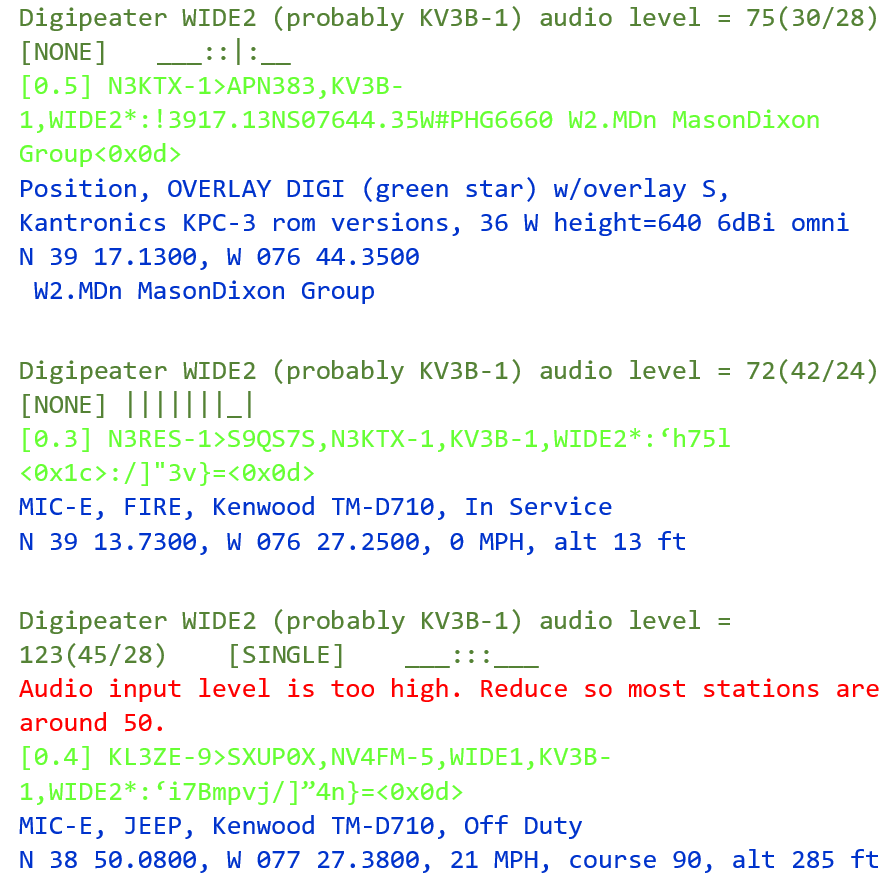}
		\caption{Raw APRS packets received by the VIPER system over the span of a few minutes during testing.}
		\label{log}
	\end{subfigure}
	\hfill
	\begin{subfigure}[b]{0.45\textwidth}
		\includegraphics[width=\linewidth]{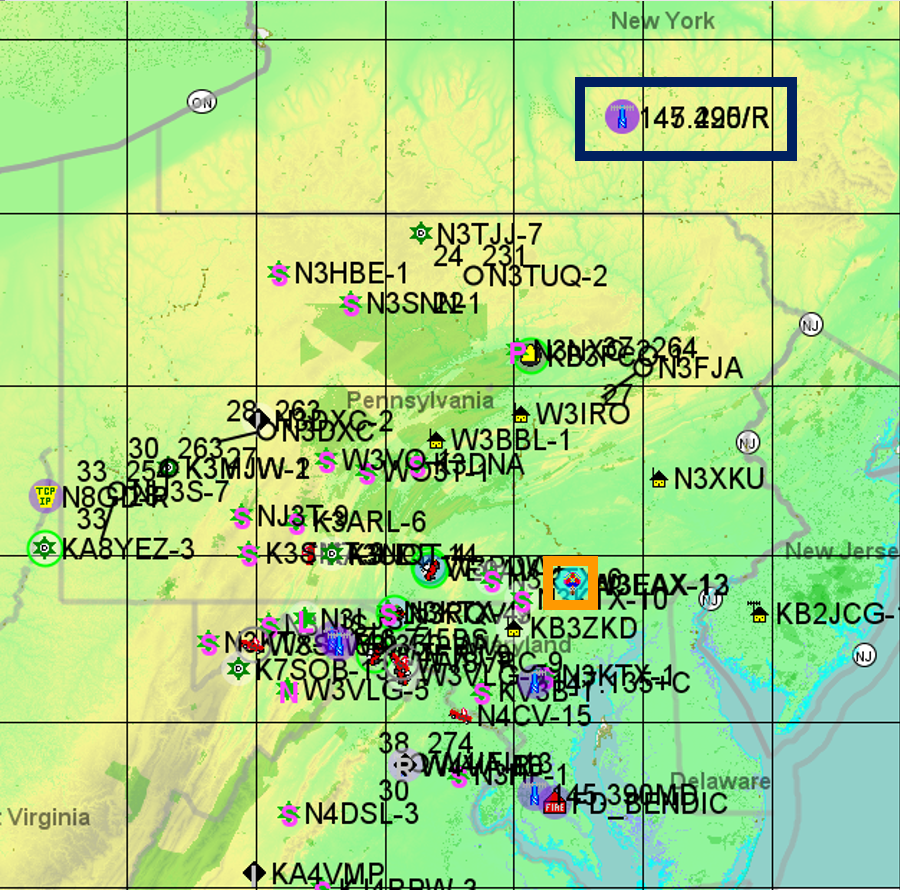}
		\caption{A map of APRS stations picked up by the VIPER system. }
		\label{ny}
	\end{subfigure}
    \hfill
    \caption{Preliminary test results of the VIPER system.}
    \label{fig3}
\end{figure}

VIPER’s ability to track a high altitude balloon was subsequently tested during the 75th balloon launch of the University of Maryland (UMD) Nearspace Program, dubbed NS-75. Using the whip antenna and Baofeng radio as an APRS receiver during the flight, the system was able to track the position of the balloon and display a live-updating map of the location up to 9 km, at which point the balloon was out of range. VIPER re-acquired the balloon’s signal upon descent, when the balloon was at an altitude of approximately 6 km. Figure 3 shows the map output generated during the flight, which was updated as new packets were received. The two balloon icons, labeled W3EAX-12 and W3EAX-13, indicate the GPS coordinates from the last received packets from the balloon’s two HABduinos, and an annotation next to each icon indicates the age of the most recently received packet. Although VIPER stopped receiving packets when the balloon ascended to an altitude that was out of range of the whip antenna, it was able to acquire packets during ascent and bridged the gap on the map with a straight line.

\begin{figure}[h]
	\begin{subfigure}[b]{0.3\textwidth}
		\includegraphics[width=\linewidth]{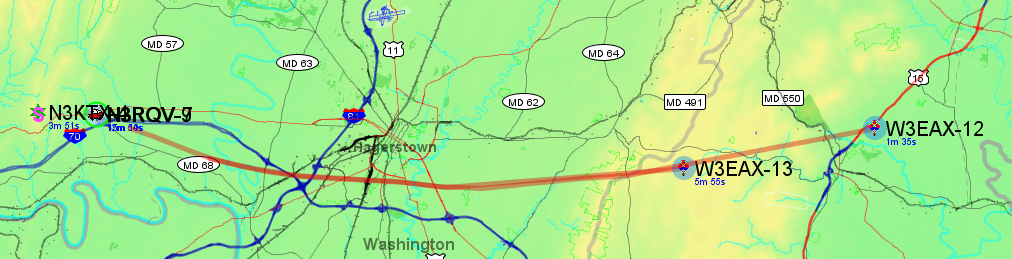}
		\caption{A screenshot taken a few minutes into flight, showing the balloon’s path (marked in red).}
		\label{path}
	\end{subfigure}
    \hfill
    \begin{subfigure}[b]{0.3\textwidth}
		\includegraphics[width=\linewidth]{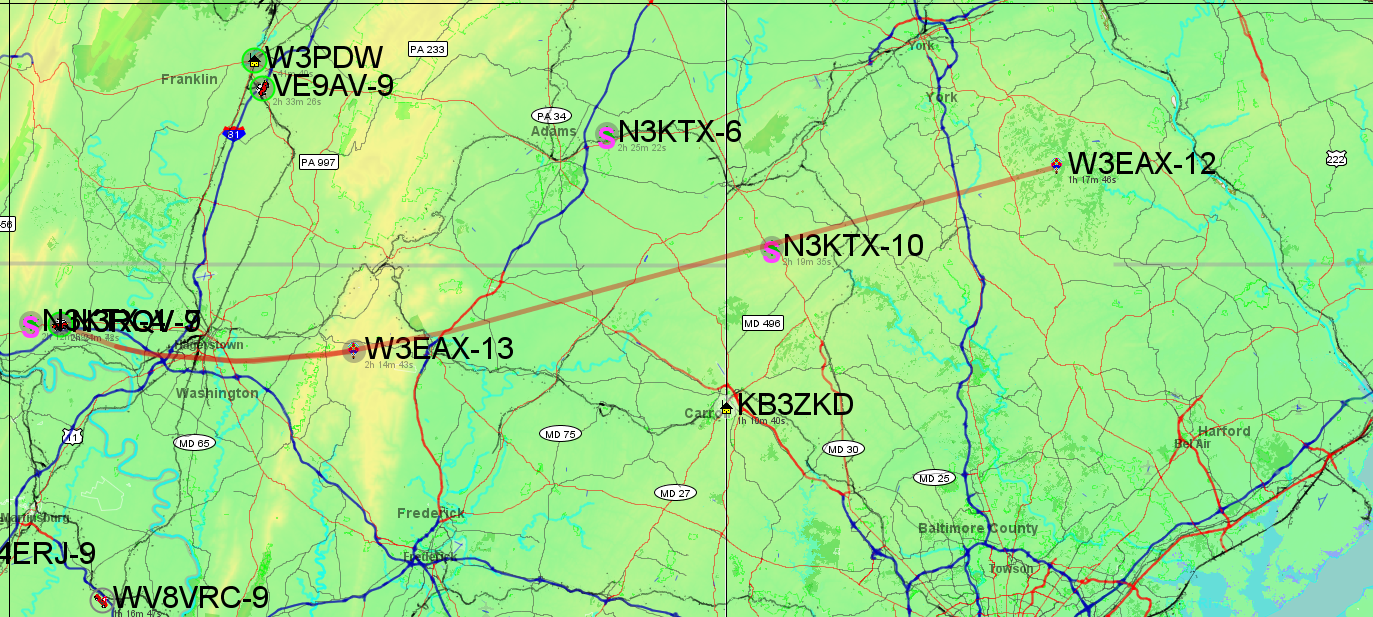}
		\caption{A screenshot taken at the end of the flight, showing the final landing location of the HAB.}
		\label{endofflight}
	\end{subfigure}
    \hfill
    \begin{subfigure}[b]{0.3\textwidth}
	\includegraphics[width=\linewidth]{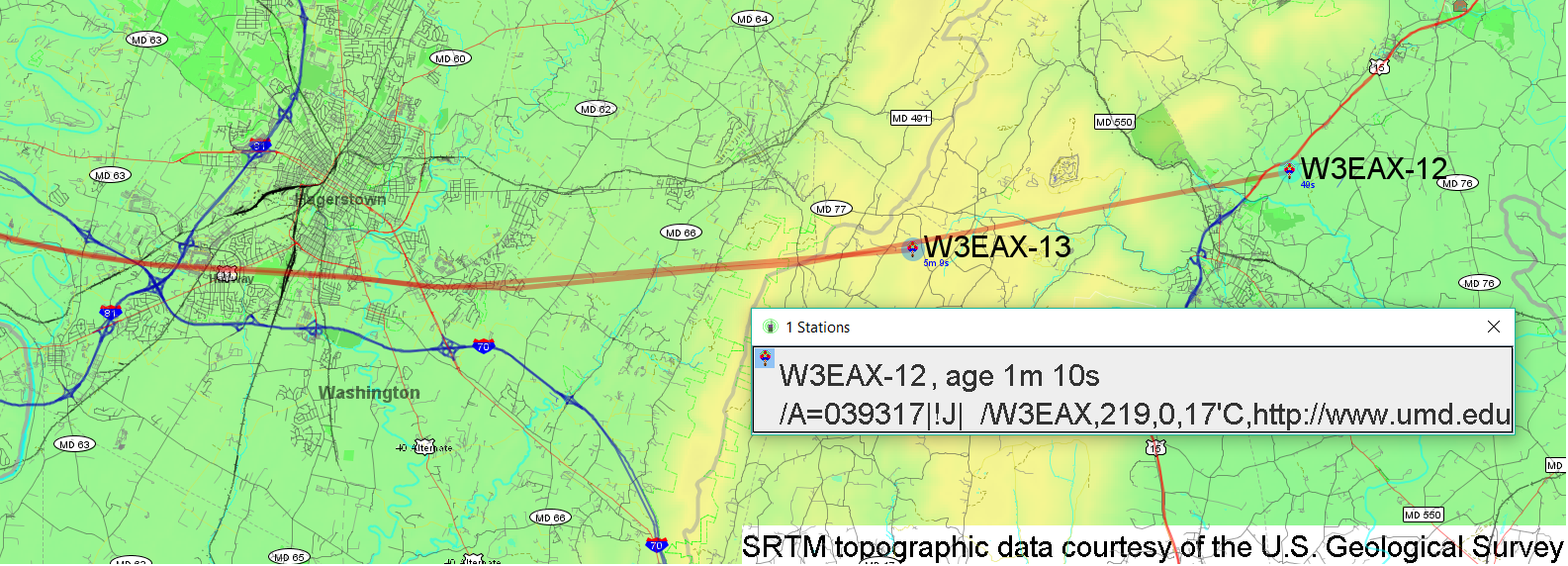}
	\caption{A close-up of the map display VIPER outputted during the NS-75 flight.}
	\label{closeup}
	\end{subfigure}
	\hfill
    \caption{The map output generated during the NS-75 flight.}
    \label{fig4}
\end{figure}

The final feature of VIPER to be tested, its concurrent mapping capability, was later confirmed during UMD’s 77th balloon launch (NS-77) by initially tracking the balloon prior to launch. The typical map display is shown in Figure 4 as a screengrab, with the module’s location while walking around the launch site as well as the balloon’s location clearly visible. 

\begin{figure} [h]
	\includegraphics[width=\linewidth]{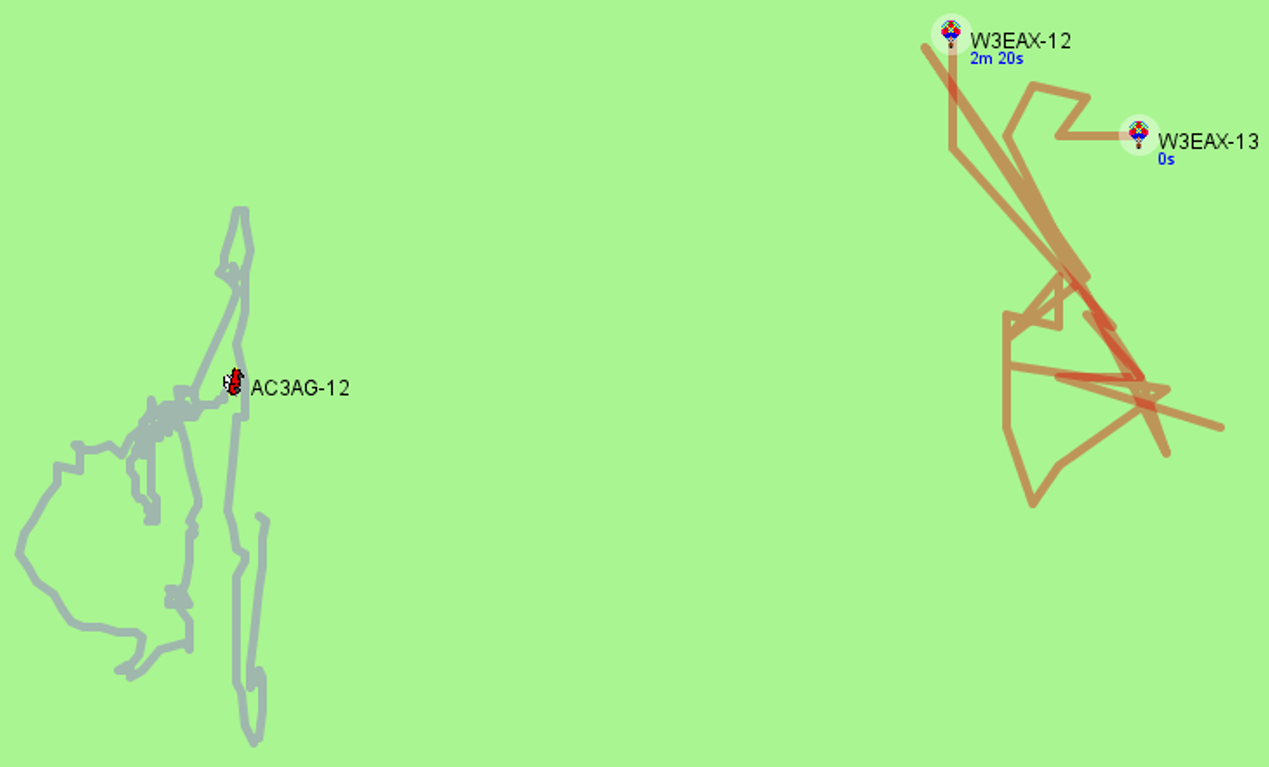}
	\hfill
    \caption{\textbf{The map output generated during the NS-77 launch. }}
    \label{fig5}
\end{figure}

\subsection{Applications of the VIPER tracking system}

Apart from the more self-evident applications of low-cost balloon tracking discussed in the introduction, the VIPER system also allows for a real-time comparison of a HAB's GPS position with that of a chase vehicle or on-foot recovery team. This GPS data, combined with altitude data that is provided by most GPS modules, can be allowed to further expand this comparison into three-dimensional space \cite{prodoningrum2015antenna}.

\begin{figure}[h]
	\includegraphics[scale=0.5]{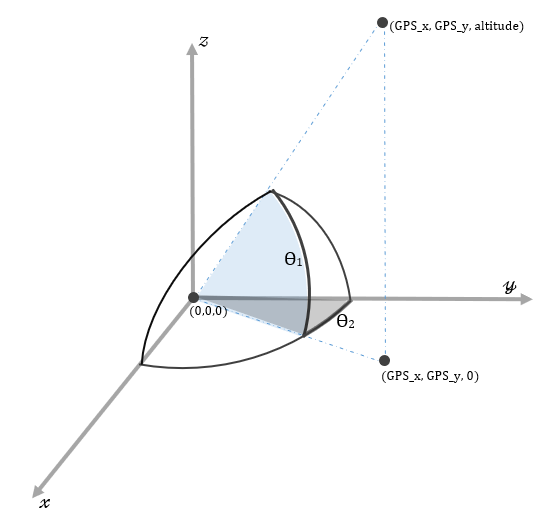}
	\hfill
    \centering
    \caption{\textbf{A diagram illustrating how azimuth angle and elevation angle can be derived from GPS and altitude data.}}
    \label{fig6}
\end{figure}

For tracking systems dependent on directional antennas (e.g. Yagi or parabolic antennas), being able to visually compare one's current location with the location of the balloon being tracked can inform either hand-held antenna pointing (when on foot) or possibly provide parameters for determining the ideal azimuth angle and elevation angle for an automated algorithm-guided antenna-pointing system. (With regards to the latter, the Adafruit GPS module employed by VIPER already provides an “angle” parameter that specifies the current absolute bearing of the module in degrees, which would assist in orienting an antenna mounted on or carried by a moving vehicle. Figure 5 illustrates the spatial relationship between the position of a tracking system (with GPS coordinates standardized to (0,0,0,)), the two-dimensional GPS coordinates of the balloon being tracked, and the three-dimensional GPS coordinates of the HAB. The elevation angle $\theta_1$ and azimuth angle $\theta_2$ of the tracking antenna can be altered to increase received signal strength. 

As Figure 6 illustrates, there is a precipitous drop in effective antenna gain for directional antennas that are oriented away from their target. For the 11.3m long, 17.39 dBi KF6A-144-18 Yagi antenna modeled in Figure 6 (a 144 MHz long boom Yagi antenna, as might be used in the tracking of a HAB with an APRS-based communications system a deviation from the line of sight to the target by as little as 15 degrees can result in a nearly 20\% reduction in gain. Given that HAB chase teams typically lose visual of their balloons within a few minutes of launch, it is not unreasonable to assume that such deviation is a common occurrence since a lack of visual feedback regarding the bearing of the antenna can detrimentally affect its positioning when done manually. A system such as VIPER would easily facilitate on-the-spot correction algorithms that could re-orient antennas (or at least provide feedback to the tracking team) so as to decrease the number of lost packets and increase the range of radio communication with the balloon.

\begin{figure}
	\begin{subfigure}[b]{0.45\textwidth}
		\includegraphics[scale=0.35]{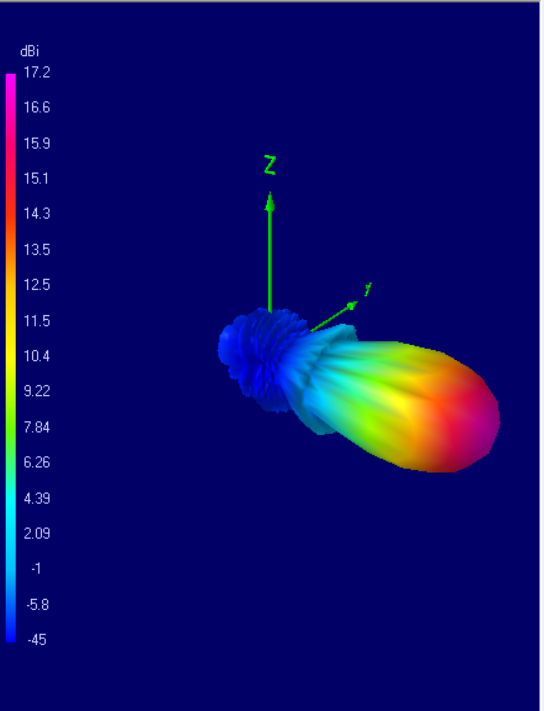}
        \centering
		\caption{Three dimensional rendering of the radiation pattern of a Yagi antenna.}
		\label{log}
	\end{subfigure}
	\hfill
	\begin{subfigure}[b]{0.45\textwidth}
		\includegraphics[width=\linewidth]{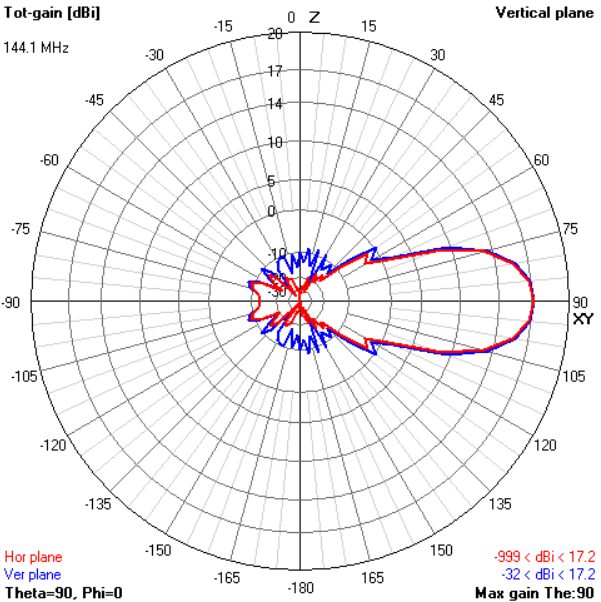}
		\caption{Two dimensional plot showing both the vertical and horizontal planes of a Yagi antenna.}
		\label{ny}
	\end{subfigure}
    \hfill
    \caption{The radiation pattern of a 144 MHz long boom Yagi antenna. Plots made using 4nec2 \protect\cite{4nec2}; NEC file for the KF6A-144-18 Yagi antenna courtesy of Roger Cox\protect\cite{cox}.}
    \label{fig7}
\end{figure}

\section{Conclusion}
There is a clear need in the ballooning community for a low-cost, versatile, handheld tracking system that can be readily used for on-foot or vehicle-based tracking. For APRS-based systems in particular, the need for an adaptable system is paramount as the limitations of a line-of-sight, VHF based radio communications system dictate that a tracking rig be optimized for maximum gain as opposed to convenience. 

The proposed system, VIPER, has been already prototyped and undergone preliminary field testing, is sufficiently customizable as to allow for HAB groups to substitute their own equipment or other off-the-shelf parts and tweak the design for their own use. It also allows for teams to monitor their position and compare it against the position of the balloon via a live-updating map, thereby simplifying the tracking process and making it easier to plan chase routes, etc. Furthermore, the ability to read in APRS packets and GPS coordinates in real time on a hand-held system promises new avenues of development in terms of mobile antenna-orienting systems and other types of positionally-aware tracking hardware.

\bibliography{biblio}

\end{document}